\documentclass[aps,prb,twocolumn,superscriptaddress]{revtex4-2}

\usepackage{graphicx}   
\usepackage{amsmath}    
\usepackage{amssymb}    
\usepackage{hyperref}   
\usepackage{xcolor}     
\usepackage{bm}
\usepackage{wasysym}
\usepackage{float}
\usepackage{lineno}

\begin{document}
	
\title{Fine Structures of Berry Curvature and Unquantized Valley Chern Numbers in Valley Photonic Crystals}

\author{Wei Dai}
\thanks{These authors contributed equally to this work.}
\affiliation{Department of Physics, Institute of Science Tokyo, 2-12-1 Ookayama, Meguro-ku, Tokyo 152-8550, Japan}

\author{Taiki Yoda}
\thanks{These authors contributed equally to this work.}
\affiliation{Department of Physics, Institute of Science Tokyo, 2-12-1 Ookayama, Meguro-ku, Tokyo 152-8550, Japan}

\author{Yuto Moritake}
\affiliation{Department of Physics, Institute of Science Tokyo, 2-12-1 Ookayama, Meguro-ku, Tokyo 152-8550, Japan}
\affiliation{Institute of Industrial Science, The University of Tokyo, 4-6-1 Komaba, Meguro-ku, Tokyo 153-8505, Japan}

\author{Masaya Notomi}
\email{Contact author: notomi@phys.sci.isct.ac.jp} 

\affiliation{Department of Physics, Institute of Science Tokyo, 2-12-1 Ookayama, Meguro-ku, Tokyo 152-8550, Japan}
\affiliation{NTT Basic Research Laboratories, NTT Corporation, 3-1 Moriosato-Wakamiya, Atsugi, 243-0198, Japan}
\affiliation{Nanophotonics Center, NTT Corporation, 3-1 Moriosato-Wakamiya,
	Atsugi, 243-0198, Japan.}

\date{\today}
\begin{abstract}
	Valley photonics has emerged as a promising platform in topological photonic systems, yet the topological nature of valley-dependent phenomena remains unsettled. Theoretically, inter-valley scattering may occur with structural imperfections, and global Chern numbers vanish due to time-reversal symmetry. As a result, valley-dependent topology is locally defined around $K(K')$ points in the half-Brillouin zone (HBZ). While half-integer valley Chern numbers have been widely assumed, their quantization and topological validity remain controversial. Here, we systematically investigate a continuous spectrum of valley photonic crystal designs by evaluating their Berry curvatures, valley Chern numbers, and angular momenta. We show that valley Chern numbers are generically unquantized and instead form a continuous spectrum varying with structural parameters. We further reveal previously unexplored fine structures in the Berry curvature distribution in momentum space. The unquantized valley Chern numbers are attributed to inter- and intra-valley cancellation of Berry curvature, highlighting the absence of a protecting mechanism for quantization. Our results call for a reassessment of valley-dependent topology and provide a more rigorous framework for interpreting valley-related photonic phenomena.
	
\end{abstract}

\maketitle

\section{Introduction}

Inspired by electronic two-dimensional Dirac materials \cite{Xiao.D-valley,W.Yao_PhysRevB.77.235406}, the photonic analog of valleytronics has led to novel light manipulation via the valley degree of freedom as a new degree of freedom. Similar to two-dimensional Dirac materials such as graphene\cite{CN_graphene} and monolayer transition metal dichalcogenide\cite{XiaoD_coupledspin}, two-dimensional photonic crystals (PhC) with hexagonal lattice structure has $K/K'$ valleys at the corner of the hexagonal Brillouin zone. Breaking two-fold rotational symmetry around the $z$ axis ($C_{2z}$ symmetry) causes opposite chirality in modes at the $K$ and $K'$ points, i.e. the chirality of modes couples to the valley degree of freedom. The two $K$ and $K'$ valleys thus can be distinguished in hexagonal lattice PhCs without $C_{2z}$ symmetry. These PhC are sometimes called valley photonic crystals (VPhCs). The first valley-photonic phenomena proposed is the valley-Hall effect of light \cite{Onda_Hall_light,Onoda_Geometrical,CZhang_Geometric}. Following works also revealed that gapped valley states at the K and K’ points carry the characteristic angular momentum: the vortex of the Poynting vector, the in-plane circular polarization, and phase singularity of the electromagnetic fields \cite{Onoda_spinningblock,valley11_Ma2016_s0,valley3_ChenXD2017_h23} and Berry curvature\cite{valley9_Mikhail2019_h23,valley4_He2019_h16,h56_gao}. Due to the time-reversal-symmetry (TRS), opposite valleys have opposite Berry curvatures:
$\Omega_{n,-\textbf{k}}=-\Omega_{n,\textbf{k}}$, where $n$ is the band index and $\textbf{k}$ is the wavevector. This leads to a zero net Chern number in the first Brillouin zone of VPhCs. To describe the local band topology hosted by each valley, a valley Chern number is defined on half Brillouin zone (HBZ) centered at $K/K'$ points: $C_{v}^{K/K^{\prime}}=\frac{1}{2\pi} \int_{HBZ} \Omega^{K/K^{\prime}}(\mathbf{k})d^2k$. A valley PhC waveguide (VPhCWG) is constructed by connecting two spatially-inverted valley photonic crystals, which have opposite $C_v$, with a proper domain wall. Valley-polarized states localized at the domain-wall are expected to appear within the bulk band gap. The robust transport of valley interface states hs been widely observed in various types of VPhCWGs\cite{valley11_H13_Yang2020,valley1_Majiawen2019_h23,valley3_ChenXD2017_h23,valley8_h13_Kumar2022,valley4_He2019_h16}.

While most previous studies report quantized values of the valley Chern numbers: $C_{v}^{K/K'}= \pm0.5$, some have argued that $C_{v}$ is not quantized ($C_{v}<0.5$)\cite{valley4_He2019_h16} and vary continuously with the structural parameters\cite{glide1_YangJK2021_c16}. It is commonly believed that if $C_v$ is non-zero, valley-dependent topology exists and various valley-dependent phenomena can be observed. However, inter-valley scattering or pseudospin flipping is not strictly prohibited even for quantized (half-integer) $C_v$, let alone smaller values. It remains unclear how one should define the valley-dependent topology when $C_v$ can be an arbitrary value smaller than 0.5, and how the specific value of $C_v$ affects the valley-dependent properties. These questions are further underscored by recent studies that doubted and challenged the topological protection of valley interface states in both disordered waveguides\cite{disorder1_Arregui,disroder2_Rosiek2023} and sharply bent waveguides\cite{Dai2025}.

Practically, large photonic band gaps (PBGs) are usually necessary for PhC devices. In PhC waveguides, complete band gap is required to confine light in one dimension. In PhC nanocavity lasers, large PBG supports high-Q, sharp resonance peaks and reduces device size and power threshold\cite{pbg_Akahane2003,pbg_chem_detec_laser,pbg_edriven_laser}. A commonly used design of VPhCs is honeycomb arrays of air holes in dielectric slabs, which offers large band gap in optical frequency. Applications such as compact lasers\cite{valley_laser_Gong2020,valley_laser_Liu:22} and valley routing circuits\cite{valley1_Majiawen2019_h23,T13_valley_filter} have been demonstrated with this design. In the hole-type VPhCs, the lowest two TE bands form the conical dispersion (Dirac cone) at the two K points. When the $C_{2z}$ symmetry of the photonic crystal is broken, the degenerate two states at the K points split and the gapped states exhibit the valley-dependent phenomenon. Triangular lattice designs are also reported as a special case of honeycomb-lattice with one sublattice vanishing. It is widely considered that the valley-dependent properties in both cases arise from the lifted Dirac cone, which is analogous to inversion-asymmetric graphene \cite{Xiao.D-valley,W.Yao_PhysRevB.77.235406}. However, the treatment of triangular lattice VPhCs as a special case of honeycomb-lattice is not rigorous. The base point of $C_{2z}$ symmetry, whose breaking gives rise to the valley-dependent properties, is different in a triangular and honeycomb lattice. There lacks fair, systematic evaluation of the valley-dependent properties in a triangular-lattice and honeycomb-lattice PhCs, and the consequences of such difference in symmetry is unclear.

In this paper, we first propose a generic scheme to build valley PhCs based on continuous structural deformation. We investigate the valley-dependent properties for the first three bands, especially focusing on the evolution of Berry curvature and angular momentum at $K$ point as the structural parameters change. We demonstrate that Dirac cone is not a necessary condition for Berry curvatures to emerge. We also show that from a practical perspective, where large PBG is necessary, the triangular-lattice design have no essential difference from the honeycomb-lattice designs regarding valley-dependent properties. We also demonstrate with several examples that the valley Chern number is generally not quantized in VPhCs.

\section{Calculation method}

\subsection{Structural model}
\begin{figure*}[hbt!]
	
	\centering
	\includegraphics[width=1.0\textwidth]{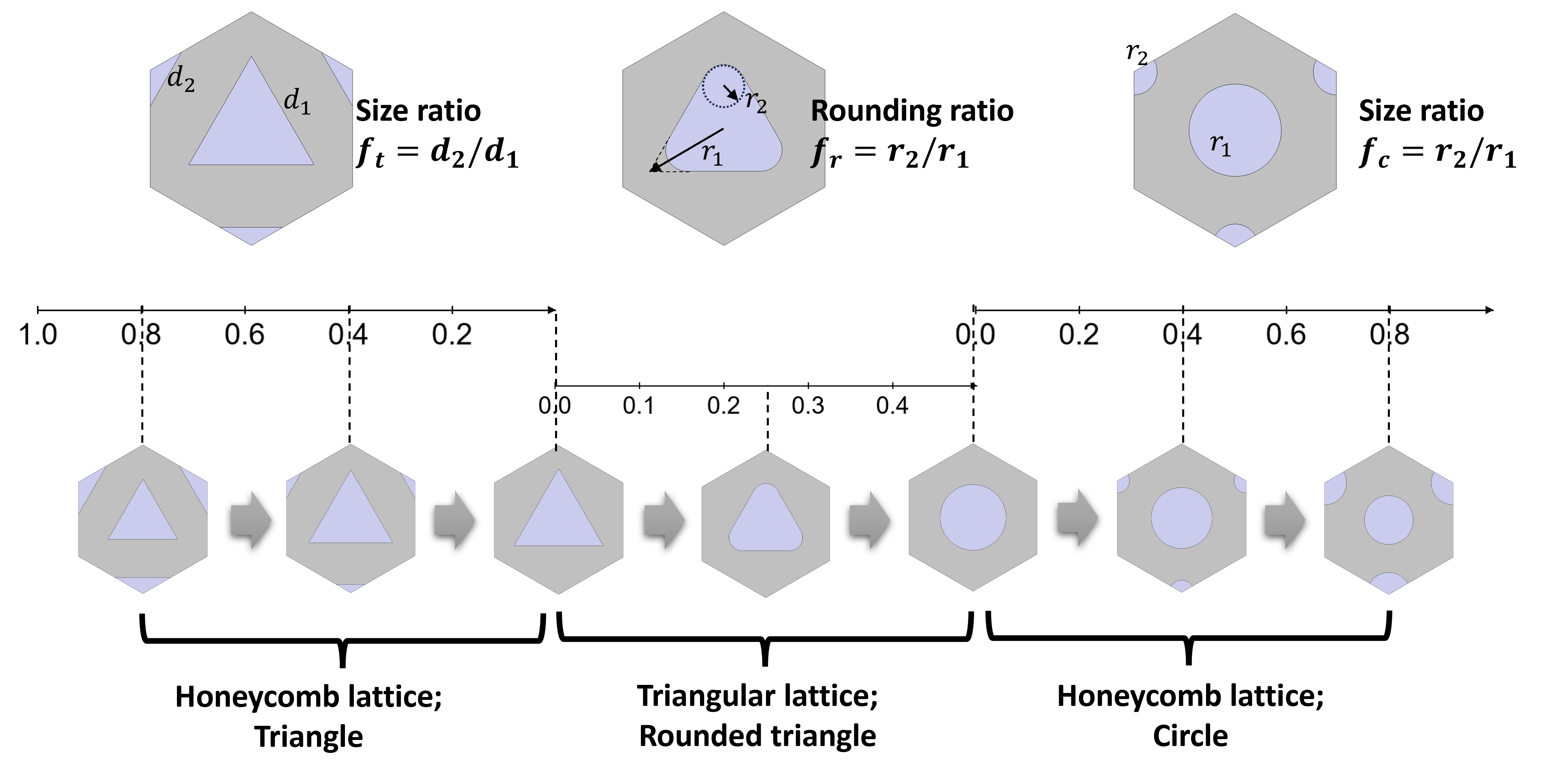}
	\caption{Illustration of structure model. Starting from a honeycomb lattice, the investigated photonic crystal hole size and shape are gradually modified, to a triangular-lattice and then back to a honeycomb-lattice. The rounding factor $f_r$ defines the rounded triangle shape in the triangular-lattice. The radius ratios $f_t$ and $f_c$ define the perturbation strength in the honeycomb lattice. $f_r$ varies in range [0,0.5] and $f_{t/c}$ varies in range [0,1]}

\end{figure*}

We investigate PhCs with honeycomb or triangular arrays of air holes in a dielectric slab.
 We consider a continuous evolution of lattice shapes in a hole-type silicon PhC. Figure1 shows the Wigner-Seitz unit cells of some typical designs. The hole shapes in the honeycomb-lattice is either circular or triangular. We call them Cir-HPC and Tri-HPC respectively. The honeycomb-lattice becomes a triangular-lattice when the air holes in one sublattice is shrunken to zero. One can further round the vertices of the triangle to control the $C_{2z}$ symmetry. As shown in the top middle image in Fig.1, the rounded triangle is constructed by replacing three corners with arcs of a circle while keeping the shape smooth. The limit of such rounding is where the triangular shape becomes circular, which restores the $C_{2z}$ symmetry. We call the triangular-lattice PhC with circular hole a Cir-TPC, and the other triangular lattice PhCs we call Tri-TPCs, where the air holes are either rounded or normal triangles. A similar shape perturbation has been reported for a pillar-type TPC by Ma et al.\cite{valley11_Ma2016_s0}, where the perturbation opens Dirac cone between second and third TE band. Here the hole-type TPC also lifts a Dirac degeneracy between band-2 and -3 during the perturbation. However, we will show that the first band gap in our TPC, which does not emerge from the perturbation and not related to the Dirac degeneracy, still hosts valley topological properties. Starting from a Tri-HPC, the PhC shape can be evolved continuously to a Tri-TPC, Cir-TPC, and finally to a Cir-HPC. We define three parameters to describe this process: Tri-HPC size ratio $f_t$, TPC rounding ratio $f_r$ and Cir-HPC size ratio $f_c$. The size ratios are defined as \( f_{t/c}=r_{t/c}^{small}/r_{t/c}^{large} \) where $r^{small}$ and $r^{large}$ are the radii of smaller and larger holes in the HPC. The rounding ratio is defined as \( f_r=r^{cir}/r^{tri} \), where $r^{cir}$ is the radius of the arcs that replace the corners of a triangular hole. By this definition, \( f_{t/c} \in [0,1] \) and \( f_r \in [0,0.5] \). The lattice constant $a$ and the area of the air hole region are kept the same for all PhCs.

\subsection{Berry Curvature calculation}

We calculate the Berry curvature numerically using Wilson-loop method. The periodicity of a system ensures that the electromagnetic wave takes the form of the Bloch wavefunction \(\textbf{E}(\textbf{r})=\text{exp}(-i\textbf{k}\cdot\textbf{r})\textbf{u}^{E}(\textbf{r})\) and \(\textbf{H}(\textbf{r})=\text{exp}(-i\textbf{k}\cdot\textbf{r})\textbf{u}^{H}(\textbf{r})\), where \(\textbf{u}^{E,H}\) is the periodic part of the Bloch function.
The Maxwell equations for \(\textbf{u}^{E,H}\) can be recast as the generalized eigenvalue problem\cite{PhysRevLett.100.013904,PhysRevA.78.033834,PhysRevA.81.053803,PhysRevLett.104.087401,lu2014topological,PhysRevB.92.125153}:
\begin{gather}
	A_{\textbf{k}}\textbf{u}_{n\textbf{k}}=\omega_{n\textbf{k}}B\textbf{u}_{n\textbf{k}},
	\label{eq:Maxwell}
	\\
	A_{\textbf{k}} =
	\begin{pmatrix}
		0 && -ic(\nabla-i\textbf{k})\times
		\\
		ic(\nabla-i\textbf{k})\times && 0
	\end{pmatrix}
	, \ B=
	\begin{pmatrix}
		\varepsilon(\textbf{r}) && 0
		\\
		0 && 1
	\end{pmatrix}
	\\
	\textbf{u}_{n\textbf{k}} = (\sqrt{\varepsilon_{0}}\textbf{u}^{E}_{n\textbf{k}}, \sqrt{\mu_{0}}\textbf{u}^{H}_{n\textbf{k}})^{T}/2\sqrt{U_{n\textbf{k}}},
\end{gather}
where \(U_{n\textbf{k}}\) is the time-averaged electromagnetic energy stored in the unit cell.
The $z$ component of the Berry curvature is defined as\cite{PhysRevLett.100.013904,PhysRevA.78.033834,PhysRevA.81.053803,PhysRevB.92.125153} 
\begin{equation}\Omega_{n\textbf{k}}=-2\text{Im}\left\langle\frac{\partial \textbf{u}_{n\textbf{k}}}{\partial k_{x}} \biggl\lvert B \biggr\rvert \frac{\partial \textbf{u}_{n\textbf{k}}}{\partial k_{y}} \right\rangle,
\end{equation}
where the inner product represents the integral over the unit cell.
To numerically calculate the Berry curvature, we discretize the $k$-space and numerically calculate the overlap integral $M(\textbf{k}_{1},\textbf{k}_{2})$ defined as\cite{Berry_calculation,valley5_Wu2017_s0} \begin{equation}M(\textbf{k}_{1},\textbf{k}_{2})=\langle \textbf{u}_{n\textbf{k}_{1}} \lvert B \rvert \textbf{u}_{n\textbf{k}_{2}} \rangle.
\end{equation}
The Berry curvature at \(\textbf{k}\) is given by
\begin{equation}
\Omega_{n\mathbf{k}}
= -\frac{\text{Im}\left[
	M(\mathbf{k}_{1},\mathbf{k}_{2})
	M(\mathbf{k}_{2},\mathbf{k}_{3})
	M(\mathbf{k}_{3},\mathbf{k}_{4})
	M(\mathbf{k}_{4},\mathbf{k}_{1})
	\right]}{\Delta k_x\, \Delta k_y}.
\end{equation}
with
\begin{gather}
	\textbf{k}_{1} = \left(k_{x}-\frac{1}{2}\Delta k_{x}, k_{y}-\frac{1}{2}\Delta k_{y}\right),
	\\
	\textbf{k}_{2} = \left(k_{x}+\frac{1}{2}\Delta k_{x}, k_{y}-\frac{1}{2}\Delta k_{y}\right),
	\\
	\textbf{k}_{3} = \left(k_{x}+\frac{1}{2}\Delta k_{x}, k_{y}+\frac{1}{2}\Delta k_{y}\right),
	\\
	\textbf{k}_{4} = \left(k_{x}-\frac{1}{2}\Delta k_{x}, k_{y}+\frac{1}{2}\Delta k_{y}\right).
\end{gather}
We set $\Delta k_{x}a/(2\pi)$ and $\Delta k_{y}a/(2\pi)$ to $1/36$ in the numerical calculation.

\section{Results}
\subsection{Band structures and at-K Berry curvatures}
\begin{figure*}[hbt!]
	\centering
	\includegraphics[width=1.0\textwidth]{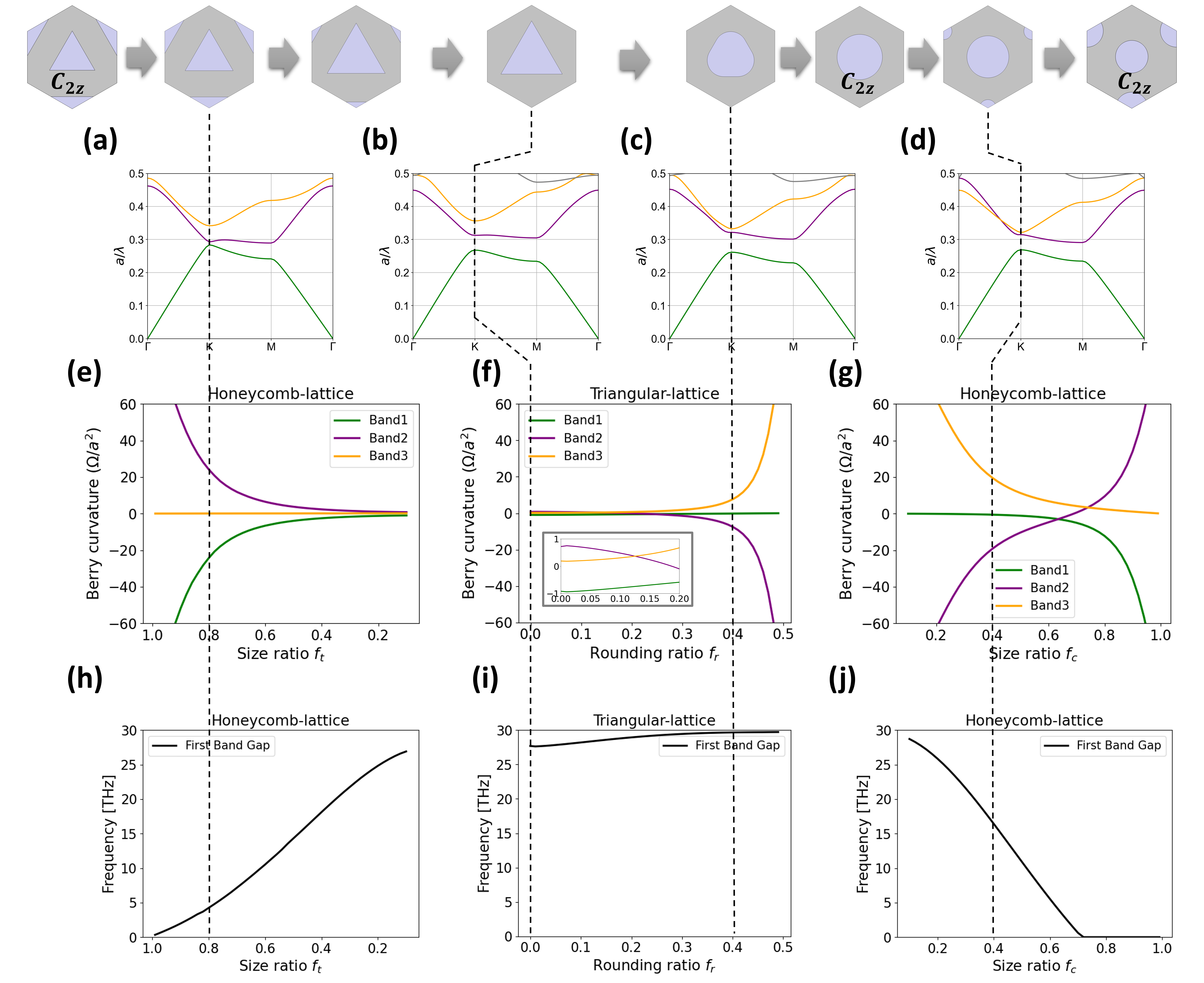}
	\caption{ Top row shows the unit cells of some investigated VPhCs. The lattice possesses $C_{2z}$ symmetry when $f_t=1.0$, $f_r=0.5$, and $f_c=1.0$ (a-d) the band diagram of four typical VPhCs, with structural factors $f_t=0.8$, $f_t=0$, $f_r=0.4$ and $f_c=0.4$ respectively. (e-g) the normalized Berry curvatures of the first three bands for varying structural factors. The first band is in green, the second band in purple and the third band in yellow. The black dashed lines are aligned to the values of structural factors. (h-j) the first band gap (in THz unit) vs. structural factor.    
	}
\end{figure*}

Now we numerically investigate the evolution of band structure in the parameter space of $f_t$,$f_r$ and $f_c$. The band structure and eigenmodes are calculated via the finite-element-method (FEM) using COMSOL. For simplicity we consider a two dimensional PhC model. The lattice constant a=400nm. The effective permittivity of silicon is set to 2.65. The filling factor of the air holes is $10.7\%$ , so that the frequency of each band does not drastically change with different shape parameters. Here we focus on the first three bands of the TE-polarization. We first calculate the at-K Berry curvatures, namely the Berry curvatures at K points ($\Omega_n^K$) for each band, which serves as a rough evaluation of the valley topology. We also focus on the band gap at K point $\omega_{ij}^K$. Although we discuss slab-type structures and TE-polarization here, the situation can be easily expanded to pillar-type structures and TM-polarization modes.

Firstly, we investigate the Tri-HPC with different hole size parameter $f_t$ ($f_r=0$ in this case). At the trivial case of $f_t=1$, the honeycomb lattice is not perturbed and preserves $C_{2z}$ symmetry. In this case band-1 and -2 are degenerate at the K(K') point. At $f_t=0.8$, as shown in Fig.2(a), the degeneracy of band-1 (green) and band-2 (purple) is slightly lifted and a tiny band gap $\Delta \omega_{21}=0.009\frac{2\pi c}{a}$ opens. Band-3 is far from band-2 at K point and $\Delta \omega_{32}=0.049\frac{2\pi c}{a}\gg \Delta \omega_{21}$. Therefore, the chirality is dominantly hosted at the first two bands. The calculated Berry curvatures are: $\Omega_1^K/a^2=-24.2$, $\Omega_2^K/a^2=24.1$ and $\Omega_3^K/a^2=0.1$ at $f_t=0.8$. $\Delta \omega_{21}$ increases and $\Delta \omega_{32}$ decreases as $f_t$ grows smaller. Fig.2(h) shows the evolution of $\Omega_n^K/a^2$ in this process. One can see that for large $f_t$, $\Omega_1^K\approx -\Omega_2^K$ and $\Omega_3^K\approx 0$. For smaller $f_t$, the influence of band-3 is no longer negligible. For example at $f_t=0.1$, $\Omega_1^K/a^2=-1.0$,  $\Omega_2^K/a^2=0.80$ and $\Omega_3^K/a^2=0.18$.

Next, we investigate the Tri-TPC with different rounding parameter $f_r$. When $f_r=0$, the air holes have a regular triangular shape. As shown in Fig.2(b), this Tri-TPC has large band gaps at K point: $\Delta \omega_{21}=0.045\frac{2\pi c}{a}$, $\Delta \omega_{32}=0.044\frac{2\pi c}{a}$. The Berry curvatures are relatively small: $\Omega_1^K/a^2=-0.92$, $\Omega_2^K/a^2=0.72$ and $\Omega_3^K/a^2=0.19$. For $f_r<0.19$, $\Omega_2^K/a^2$ is positive; band-1 and -2 are separated by a large gap but have opposite Berry curvature values, as is shown in the inset image of Fig.2(f). As $f_r$ grows larger, band-2 and -3 move closer to each other, $\Delta \omega_{21}$ increases and $\Delta \omega_{32}$ decreases, and $\Omega_1^K/a^2$ gradually approaches 0, $\Omega_2^K/a^2$ becomes smaller and $\Omega_3^K/a^2$ larger.  At $f_r=0.4$, $\Delta \omega_{21}=0.060\frac{2\pi c}{a}$ and $\Delta \omega_{32}=0.011\frac{2\pi c}{a}$. The second band gap at K point is much smaller and Berry curvature is dominantly hosted at band-2 and -3: $\Omega_1^K/a^2=-0.19$, $\Omega_2^K/a^2=-7.53$ and $\Omega_3^K/a^2=7.68$. At the limit $f_r=0.5$, the Tri-TPC becomes Cir-TPC, which has circular air holes. The Cir-TPC has $C_{2z}$ symmetry. Band-2 and -3 are degenerate at K(K') point. Berry curvatures are zero everywhere except at the degeneracies where they cannot be defined.

Finally, we investigate the Cir-HPC with different hole size parameter $f_c$ ($f_r=0.5$ in this case). As $f_c$ grows larger $\Delta \omega_{21}$ decreases and $\Delta \omega_{32}$ increases. Fig.2(g) shows the Berry curvature evolution. $\Omega_1^K/a^2$ is negative and becomes smaller. $\Omega_2^K/a^2$ is originally negative, turns positive at $f_c=0.69$ and becomes larger. $\Omega_3^K/a^2$ becomes smaller and approaches 0 as $f_c$ approaches 1.0. At the limit $f_c=1$, the honeycomb lattice is unperturbed and preserves $C_{2z}$ symmetry. Band-1 and -2 become degenerate at the K(K') point.
 
In applications, it is usually necessary to keep a large band gap. Here we focus on the first band gap, which is most commonly used, and examine the three types of PhCs. Fig.2(h-j) shows the complete first band gap $G_1$ (not $\Delta \omega_{21}$) in THz unit for the three types of PhCs. For the Tri-HPC, there exists a range for $f_t$ where both $G_1$ and $|\Omega_{1,2}^K|/a^2$ are sufficiently large. For the TPCs, although $G_1$ is always large, $|\Omega_{1,2}^K|/a^2$ is relatively small. For the Cir-HPC, although $\Omega_1^K$ and $\Omega_2^K$ have large, opposite values at $f_c>0.7$, there is no complete first band gap and therefore it would be difficult to exploit the large Berry curvatures. It is worth noting that a two-band model involving only one Dirac point may well explain $\Omega_i^K$ in the Tri-HPC, where the influence of band-3 is negligible, but cannot explain the complex evolution in Tri-TPC and Cir-HPC. In most of the large band gap regions in Tri-TPC and Cir-HPC, $\Omega_1^K$ and $\Omega_2^K$ have the same sign. Therefore it is generally not proper to treat the Tri-TPC as a special case of Tri-HPC where opening the Dirac cone creates a pair of valley states opposite in chirality in the first two bands. A three-band model is necessary to approximate our calculation results for at-K Berry curvatures. (see appendix A)
\begin{figure*}[hbt!]
	\centering
	\includegraphics[width=1.0\textwidth]{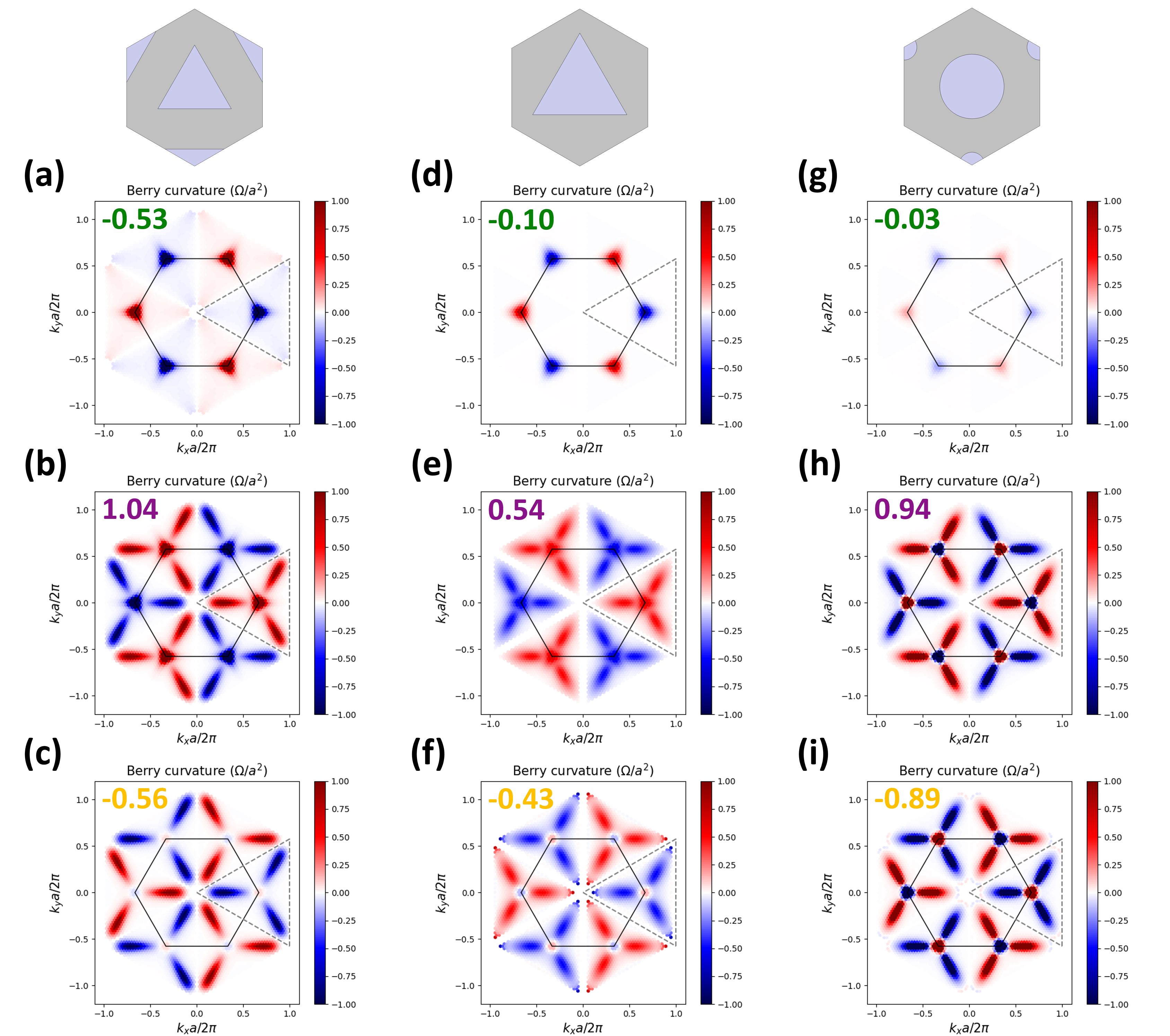}
	\caption{Berry curvature distributions for (a-c) $f_t=0.8$ Tri-HPC, (d-f) $f_r=0$ TPC and (g-i) $f_r=0.4$ TPC. The top row shows the Berry curvature for band-1, middle row for band-2 and the bottom for band-3. The corresponding unit cells are shown above each column. In each plot, the solid hexagon show the first Brillouin zone. The dashed triangles show the half-Brillouin zone centered at K point. The calculated valley Chern number are shown at the left top in each plot. The color bar shows the value of normalized Berry curvature $\Omega/a^2$.
	}
\end{figure*}

\subsection{Berry curvatures in half-Brillouin zone and valley Chern number}
In this section we investigate the Berry curvature distribution in the half-Brillouin-zone (HBZ) and the corresponding valley Chern number. In the following, we show fine structures of the Berry curvature such as "triangle", "petal"and "island" regions and discuss how they affect the valley Chern number. Fig.3 shows the color map of Berry curvatures corresponding to the structures shown in Fig.2(a-d). Because there are degeneracies at $\Gamma$ point and the HBZ edge in most bands, we set the actual calculation region slightly smaller than the HBZ. (see appendix B for detail). We start from the $f_t=0.8$ Tri-HPC. Fig.3(a,b,c) shows the distribution of $\Omega_1/a^2$, $\Omega_2/a^2$ and $\Omega_3/a^2$ respectively. The solid-line hexagon shows the first Brillouin zone (FBZ), with $\Gamma$ point at its center. The dashed-line triangle shows the HBZ with K point at the center (K-HBZ). Due to the time-reversal symmetry, $\Omega(k)=-\Omega(-k)$. The HBZ centered at K point and that centered at K' point have the same Berry curvature distribution except the sign is reverted. Now we confine our discussion within the K-HBZ. $\Omega_1$ of $f_t=0.8$ Tri-HPC (Fig.3(a)) is negative in the K-HBZ. It is well localized at the vicinity of K point, forming a "triangle" region. $\Omega_2$ of $f_t=0.8$ Tri-HPC is shown in Fig.3(b). Its feature in K-HBZ can be divided into a central "triangle" region and a three-fold "petal" region extending in the $\Gamma K$ direction. Here in Fig.3(b) we see the first example of fine structures of Berry curvature in a valley PhC, which is not simply localized at K point, but a complex configuration that extends in the whole HBZ. The "triangle" part of $\Omega_2$ is localized around K point. It constitutes a "pair" with the corresponding blue triangle of $\Omega_1$, which is natural considering the small $\Delta \omega_{21}$ at K point. The "petal" region of $\Omega_2$ extends in $\Gamma K$ direction and vanishes near the $\Gamma$ point.  Now we investigate the $\Omega_3$ of $f_t=0.8$ Tri-HPC, as shown in Fig.3(c). The  $\Omega_3$ can also be roughly divided into two regions: a positive "island" centered at K point and the negative "petal" region extending in the $\Gamma K$ direction. The "petal" region forms a pair with the  corresponding “petal” region in $\Omega_2$, with approximately opposite Berry curvatures. Interestingly, The small "island" at the center has positive Berry curvatures, that is, the same sign as the "triangle" region of $\Omega_2$. The positive value gradually decreases and turns negative away from the K point. As far as valley-photonics is concerned, only the central "triangle" or "island" regions at $\Omega_{1,2,3}$ are important. However, valley Chern number is strongly related to the fine structures. The calculated valley Chern numbers of the three bands are $C_v^1=-0.53$, $C_v^2=1.04$ and $C_v^3=-0.56$. The Chern number in the first three bands add up approximately to 0. This is apparent in the $k\cdot p$ method calculation (see supplementary A), that the total Berry curvature of the first three bands is zero everywhere in the wavevector space, given that the influence of higher bands is negligible. For band-1, the Berry curvature is well localized around $K$ point, and there is only one local extremum of Berry curvature in the K-HBZ exactly at $K$. However, off-$K$ extrema exist in band-2 and -3. For band-3, the off-$K$ values are even dominant. Despite the positive $\Omega_3$ at $K$, $C_v^3$ is negative, with the major contribution from the "petal" region close to $\Gamma$ points. In this case, the valley Chern number defined at K-HBZ does not necessarily reflect the chirality around $K$ valley.

Now we move on to the Tri-TPC with different rounding parameter $f_r$. The $f_r=0$ TPC (Fig.3(d-f)) has a similar Berry curvature distribution as the $f_t=0.8$ HPC. $\Omega_2$ also has a "petal" region that forms a pair with $\Omega_3$ and a "triangle" region that forms a pair with $\Omega_1$. However, $\Omega_2$ and $\Omega_3$ of $f_r=0$ TPC are less localized and spread out in the whole HBZ. This result shows a clear trade-off between the Berry curvature localization and the band gap. A larger band gap is accompanied by smaller, less localized $\Omega^K$ in the corresponding bands. The calculated valley Chern numbers of the three bands are $C_v^1=-0.10$, $C_v^2=0.54$ and $C_v^3=-0.43$. The conservation law still holds that $C_v^1+C_v^2+C_v^3\approx 0$. Similar to that in the $f_t=0.8$ Tri-HPC, $C_v^3$ in the Tri-TPC is mostly contributed by "petal" region, and the central  region have relatively weaker Berry curvature.

Finally, we investigate the $f_r=0.4$ TPC. Fig.3(g-i) shows the Berry curvature distribution. Since $\Delta \omega_{21}<<\Delta \omega_{32}$, chirality is dominantly hosted at band-2 and -3. $\Omega_1$ is extremely small and $C_v^1$ is only -0.03. Again, the $\Omega_2$ and $\Omega_3$ distribution consists of "petal" and "island" regions. However, at both bands, the "petal" and "island" regions have opposite Berry curvatures. The calculated valley Chern numbers are $C_v^2=0.94$, $C_v^3=-0.89$, meaning that the contribution from "petal" regions are dominant. This is natural considering that the band gap of band-2 and -3 is extremely small at the "petal" region (Fig.2(c)). For technical reasons, we do not discuss the Cir-HPCs here. (see appendix B for details.) 

In Figure.4, we plot the valley Chern number $C_v$ evolution with varying structural parameters for the first three bands. $C_v$ varies continuously with structural parameters and is not quantized in most cases. $C_v^1$ (green curves) decreases as the sublattice size ratio  $f_t$ increases in Tri-HPC and approaches zero as the rounding ratio $f_r$ increases in TPC. The behavior of $C_v^1$ is consistent with that of at-K Berry curvature $\Omega_1^K$ shown in fig.2(e,f). There is an evident trade-off between $C_v^1$ and the first band gap. For Tri-HPCs with $f_t<0.75$, $C_v^1$ is less than 0.5. For band-2 and -3, $C_v$ is not naively proportional to the at-K Berry curvature. $C_v^2$ is large for small $f_t$ Tri-HPC, and approaches 0.5 when $f_t$ is close to zero. $C_v^3$ grows from zero to 0.5 as $f_t$ varies from 1.0 to 0.8, and remains around 0.5 for smaller $f_t$ values. In the TPC, as $f_r$ increases, band-2 and -3 move closer to each other near $K$ point. $\Omega_2$ and $\Omega_3$ have large, approximately opposite values. Consequently, $C_v^2$ and $C_v^3$ become larger as $f_r$ grows, and have almost opposite values. The total $C_v$ of the first three bands (gray curves) remain nearly zero for both HPC and TPC.

\begin{figure}[hbt!]
	\centering
	\includegraphics[width=0.5\textwidth]{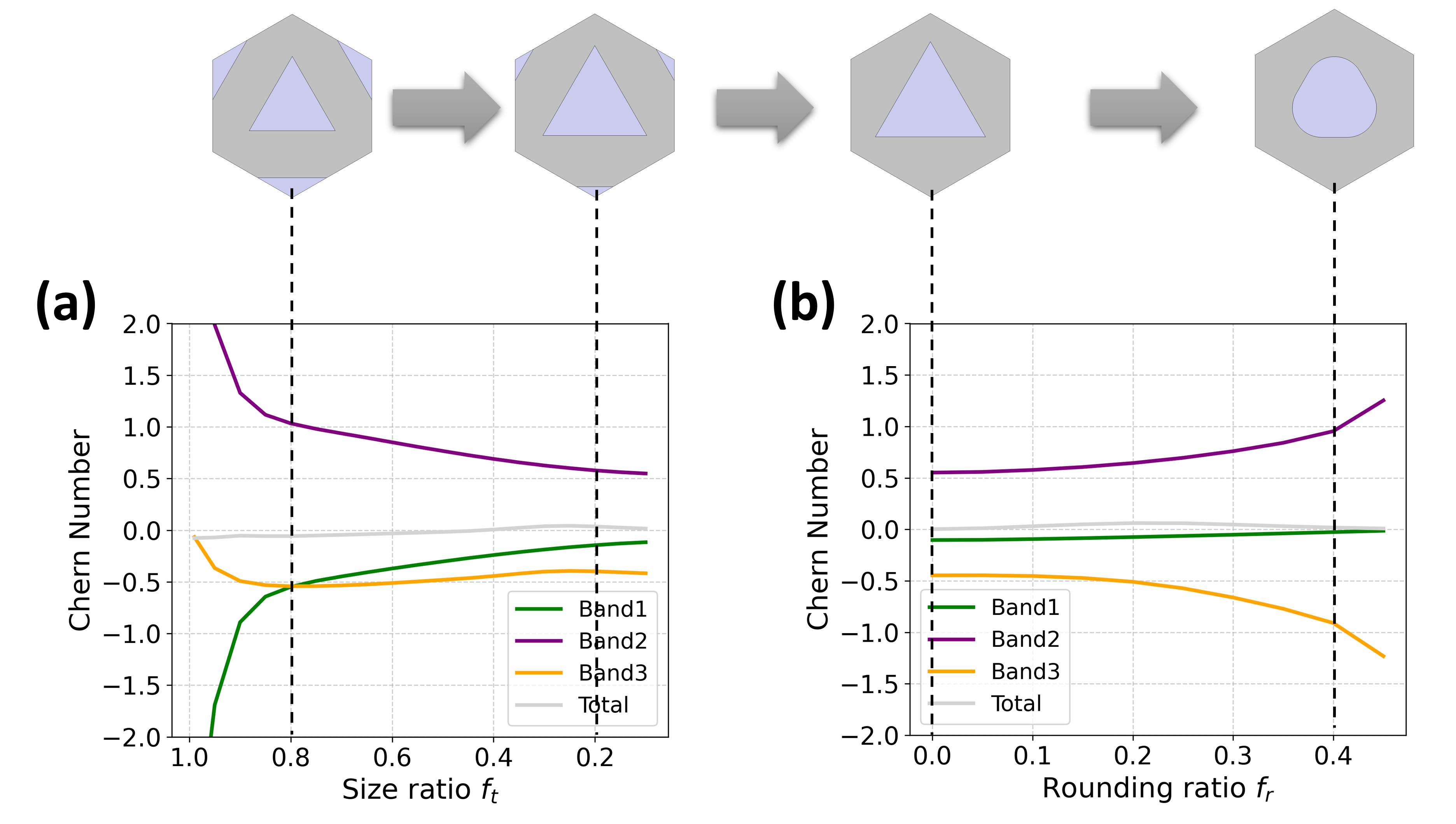}
	\caption{Valley Chern number evolution for varying structural parameters. (a) $C_v$ in the Tri-HPC. $f_t$ varies from 0.99 to 0 (b) $C_v$ in the TPC. $f_r$ varies from 0 to 0.49.
	}
\end{figure}  

To summarize, the Berry curvature distribution at band-1 is well-localized near the K(K') point at all investigated PhCs. However, the valley Chern number $C_v^1$ is not quantized in most cases. $C_v$ approaches integer or half-integer only when $\Delta \omega$ is sufficiently small. The fine structure of Berry curvature distribution at higher bands indicate that the chirality extrema is not necessarily located at $K/K'$ valleys. The off-$K$ Berry curvature extrema may have opposite sign to that of the at-$K$ extrema. When calculating the valley Chern number, cancellation not only occurs for inter-valley overlap, but also occurs in the intra-valley fine structure. Thus, valley Chern numbers are generally not quantized. Furthermore, valley Chern numbers do not necessarily reflect the local Berry curvature strength. The Berry curvature distribution and valley Chern number of TPCs and HPCs constitute a continuous spectrum as structural parameters change, showing no distinct difference. It would be reasonable to group them as the same type of valley topological insulator despite the difference in lattice configuration.

\subsection{Angular momentum}
\begin{figure*}[hbt!]
	\centering
	\includegraphics[width=1.0\textwidth]{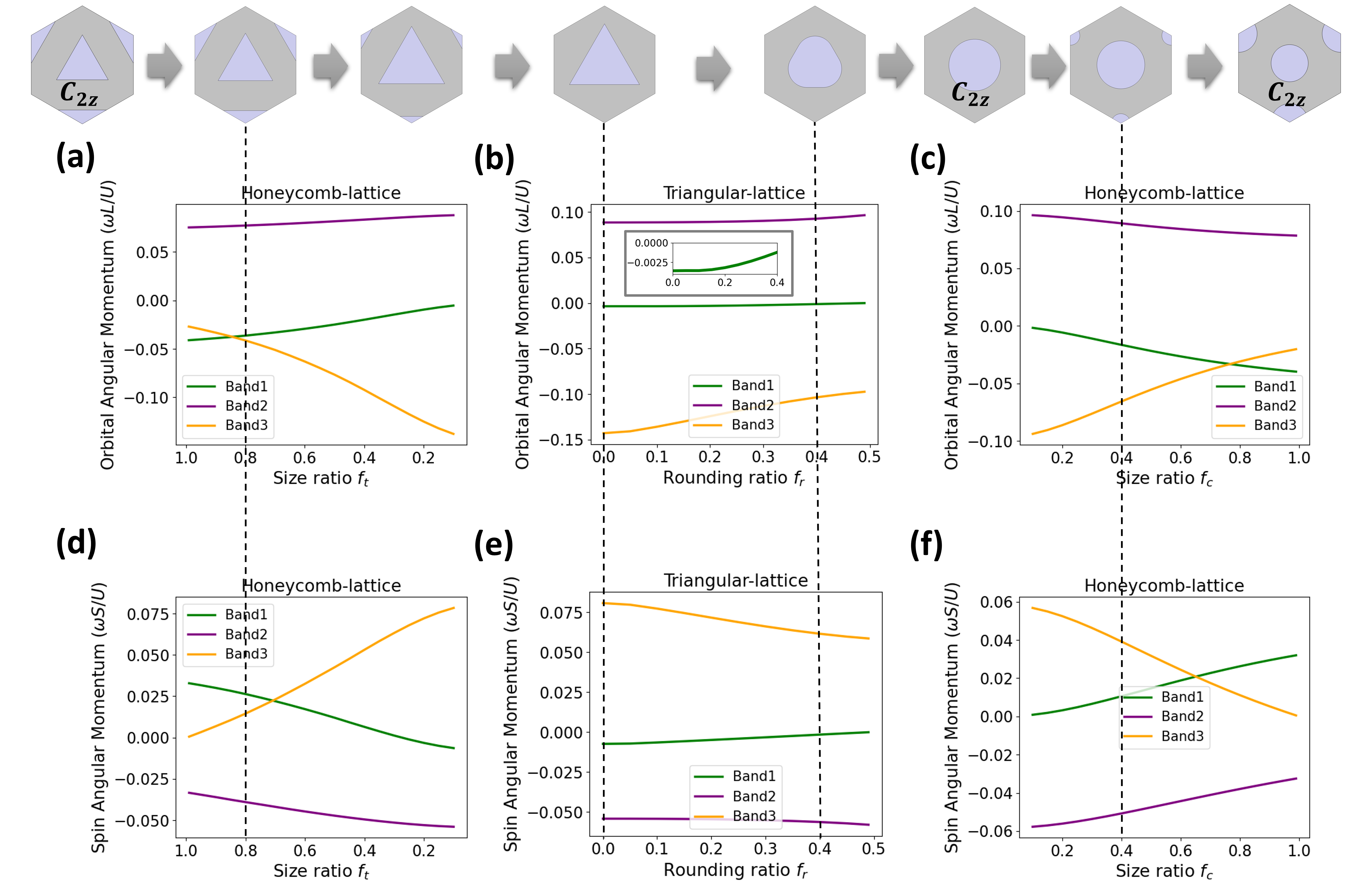}
	\caption{The calculated (a-c) orbital angular momentum and (d-f) spin angular momentum of the first three bands for varying structural factors. The first band is in green, the second band in purple and the third band in yellow. The top image shows illustrations of some VPhC unit cells. The black dashed lines cut across the corresponding structural factors (x axis) in the angular momentum plot.
	}
\end{figure*}
As a more direct demonstration of the valley-specific chirality, we calculate the $z$ component of the angular momentum in the unit cell at the $K$ point. The Abraham total angular momentum density $\boldsymbol{j}_{n \boldsymbol{k}}$ and the spin angular momentum density $\boldsymbol{j}_{n \boldsymbol{k}}^{\text {spin }}$ are defined by\cite{AM_Berry,AM_Bliokh}

\begin{gather}
	\boldsymbol{j}_{n \boldsymbol{k}}(\boldsymbol{r})=\frac{1}{c^{2}}\left(\boldsymbol{r} \times \boldsymbol{S}_{n \boldsymbol{k}}(\boldsymbol{r})\right)\\
	\boldsymbol{j}_{n \boldsymbol{k}}^{\mathrm{spin}}=\frac{1}{4 \omega} \operatorname{Im}\left[\varepsilon_{0} \boldsymbol{E}_{n \boldsymbol{k}}^{*} \times \boldsymbol{E}_{n \boldsymbol{k}}+\frac{\mu_{0}}{\varepsilon(\boldsymbol{r})} \boldsymbol{H}_{n \boldsymbol{k}}^{*} \times \boldsymbol{H}_{n \boldsymbol{k}}\right]
\end{gather}

where $\boldsymbol{S}_{n \boldsymbol{k}}=(1 / 2) \operatorname{Re}\left(\boldsymbol{E}_{n \boldsymbol{k}}^{*} \times \boldsymbol{H}_{n \boldsymbol{k}}\right)$ is the time-averaged Poynting vector. The total angular momentum (TAM) $\boldsymbol{J}_{n \boldsymbol{k}}$ and spin angular momentum (SAM) $\boldsymbol{J}_{n \boldsymbol{k}}^{\text {spin }}$ within the unit cell and are expressed as the volume integral of $\boldsymbol{j}_{n \boldsymbol{k}}$ and $\boldsymbol{j}_{n \boldsymbol{k}}^{\text {spin }}$ over the unit cell respectively. It follows from Eqs.11-12 that the Poynting vector contributes to $\boldsymbol{J}_{\boldsymbol{n} \boldsymbol{k}}$ while the in-plane electromagnetic field contributes to $\boldsymbol{J}_{\boldsymbol{n} \boldsymbol{k}}^{\text {spin }}$. The orbital angular momentum (OAM) $\boldsymbol{j}_{n \boldsymbol{k}}^{\text {orbit}}=\boldsymbol{j}_{n \boldsymbol{k}}-\boldsymbol{j}_{n \boldsymbol{k}}^{\text {spin }}$. 
The values of total and orbital angular momentum are dependent on the choice of unit cell. Like the Berry curvature calculation, we use the Wigner-Seitz cells as unit cells, and choose the central lattice sites as origins for the $\boldsymbol{j}_{n \boldsymbol{k}}^{\text {orbit}}$ calculation. 
We calculate the $z$ component of total angular momentum $J_i$, spin angular momentum $S_i$ and orbital angular momentum $L_i$ for band-i (i=1,2,3) at the K point as a function of the structural parameters. The calculated angular momentum is normalized by the angular frequency $\omega$ and the time-averaged electromagnetic energy:

\begin{equation*}
	U_{n \boldsymbol{k}}=\frac{1}{4} \int_{\text {unit cell }} d^{3} \boldsymbol{r}\left[\varepsilon_{0} \varepsilon(\boldsymbol{r})\left|\boldsymbol{E}_{n \boldsymbol{k}}\right|^{2}+\mu_{0}\left|\boldsymbol{H}_{n \boldsymbol{k}}\right|^{2}\right] \tag{3}
\end{equation*}

\begin{figure*}[hbt!]
	\centering
	\includegraphics[width=1.0\textwidth]{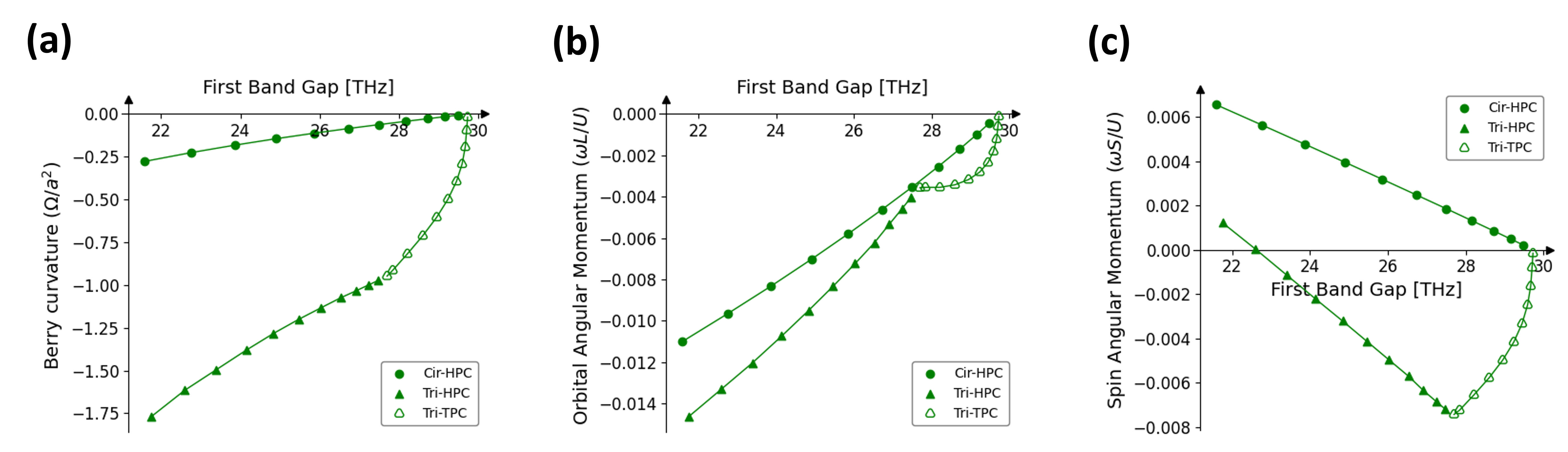}
	\caption{The (a) Berry curvature, (b) total angular momentum and (c) spin angular momentum of band-1 in three types of PhCs plot against the first band gap. Cir-HPC is in solid circular dot, Tri-HPC in solid triangle, and Tri-TPC in hollow rounded triangle.}
	
\end{figure*}

In a $f_r=0.5$ Cir-TPC, $J_1$ vanishes because the $C_{2 z}$ symmetry imposes $\boldsymbol{S}_{n,-\boldsymbol{k}}(x, y, z)=\boldsymbol{S}_{n \boldsymbol{k}}(-x,-y, z)$ and the time reversal symmetry imposes $S_{n \boldsymbol{k}}(x, y, z)=-\boldsymbol{S}_{n \boldsymbol{k}}(x, y, z)$. $J_2$ and $J_3$ cannot be determined due to the degeneracy. Similarly, for a $f_{t/c}=1$ HPC $J_1$ and $J_2$ cannot be determined and $J_3$ vanishes. Breaking the $C_{2 z}$ symmetry induces the finite total angular momentum. Note that the Wigner-Seitz unit cell does not possess $C_{2 z}$ symmetry even in equivalent sublattice ($f_{t/c}=1$) case. However, this does not affect calculation for other structures or our main conclusions. In fig.5 we omit the $C_{2 z}$-symmetric case and plot the results for $0.99\geq f_t \geq0.1$, $0\leq f_r \leq0.49$, and $0.1\leq f_c \leq0.99$. We also omit the results of $J_i$ and only show the results for $S_i$ and $L_i$ respectively.

Fig.5 (a-c) show the OAM evolution as structural parameters change. All three bands have non-zero OAM when $C_{2z}$ symmetry is broken. $|L_1|$ is generally smaller than $|L_2|$ and $|L_3|$, and $L_2$ is generally insensitive to the structural deformation. $L_1$ is extremely small in Tri-TPCs. As rounding ratio $f_r$ grows near 0.5, $C_{2z}$ symmetry is gradually restored and $L_1$ approaches zero in the Tri-TPC (inset plot in Fig.5(b)). Therefore, the TPC possesses $C_{2z}$ symmetry-dependent OAM, but the value is smaller compared to HPCs. Comparing OAM with valley Chern number shown in Fig.4, we find that for Tri-HPC and TPC, $L_i$ and $C_v^i$ behave similarly. $L_1$ and $C_v^1$ grows smaller as $f_t$ decreases and $f_r$ increases. $L_2$, $C_v^2$ stays positive; and $L_3$, $C_v^3$ stays negative in this process.
Fig.5 (d-f) show the SAM evolution as structural parameters change. Similar to OAM, $S_i$ is also non-zero when the $C_{2z}$ symmetry is broken. One may find similar dependency on structural parameter in angular momentum $L_1$, $S_1$ and Berry curvature $\Omega_1^K$. All have smaller absolute values in the Tri-TPC and grow larger as the size ratio $f_{t/c}$ become large in the HPCs. Overall, the angular momentum is not strongly correlated with at-K Berry curvature shown in Fig.2, which is sensitive to band gaps. 

 However, as noted in section.4, not all valley PhC designs are feasible considering application. If one requires a large full band gap, the large $f_{t/c}$ HPCs are not suitable even though they have larger Berry curvature and SAM. To better evaluate valley-dependent chirality, we plot the $\Omega_1^K$, $L_1$ and $S_1$ against the first band gap $G_1$ for the three types of PhCs in figure.6.

Fig.6(a) shows $\Omega_1^K/a^2$-$G_1$ relation. Here we are interested in large $G_1$ designs and focus on the data with $G_1>20 THz$. As already shown in Fig.2(h-j), this condition is satisfied by all TPCs, but only by a small portion of investigated HPCs, approximately those with size ratio smaller than 0.3. In Fig.6(a), the Tri-HPC (solid triangle marks) and Tri-TPC (hollow rounded marks) results connect smoothly and show a clear trade-off between Berry curvature and band gap. The Cir-HPC has much smaller $\Omega_1^K$ compared to other structures. Considering $\Omega_1^K$, the optimal valley PhC design falls between a Tri-HPC with small $f_t$ (large sublattice contrast) and a Tri-TPC with small $f_r$ (slightly or not rounded triangular hole). Fig.6(b) shows $L_1$-$G_1$ relation. The Cir-HPCs and Tri-HPCs have very similar, almost linear $L_1$-$G_1$ relation, and show a clear trade-off between their absolute values. Although the overall $L_1$ is small, the Tri-TPCs have larger $L_1$ given the same band gap. Comparing Fig.6 (a,b) one can observe a similar trend in OAM and Berry curvature. Fig.6(c) shows $S_1$-$G_1$ relation. Note that most Tri-TPCs and Tri-HPCs have negative $S_1$ while the Cir-HPCs have positive $S_1$ in this plot. The overall $|S_1|$ in three types of valley PhCs are the same order of magnitude. Again, given the same $S_1$, the Tri-TPCs are most advantageous for they have larger band gap. We conclude from fig.6 that both Tri-HPCs and small $f_r$ Tri-TPCs are good valley PhCs that offer both large photonic band gap and relatively strong valley-dependent properties. If one could compromise band gap to some extent, the optimal design that maximizes $\Omega_1^K/a^2$ and $L_1$ would be a Tri-HPC, while the Tri-TPC still is the best choice that maximizes the spin angular momentum $S_1$ in all investigated designs.

\section{Discussion and conclusions}
In summary, we have examined the Berry curvature and angular momentum for various types of valley PhCs in a continuous geometric spectrum. We calculated Berry curvature in half-Brillouin zone using the Wilson-loop method. The localization and absolute value of Berry curvature around $K/K'$ point shows an evident trade-off with the band gap. Berry curvature distribution at higher bands generally show fine structures along the $\Gamma-K$ direction. Most of these fine structures contain off-$K$ extrema, sometimes even with opposite signs to the at-K Berry curvature. For Berry curvature distribution localized at K point, with no significant fine structure, the resulting valley Chern number $C_v$ remains closely correlated with the local valley chirality. In such cases, inter-valley and intra-valley interferences are relatively weak. Although not strictly quantized, $C_v$ could still serve as a useful indicator of valley-selective response, and may be associated with enhanced robustness against certain types of perturbations. However, for Berry curvature exhibiting pronounced fine structures, both inter-valley and intra-valley interferences are strong. The resulting $C_v$ no longer provides a direct measure of the chirality localized near the valley. Therefore, the magnitude of $C_v$ cannot, in general, be used as a reliable indicator of valley-dependent properties. In the parametric evolution, $C_v$ varies continuously and approximates half-integer only accidentally or when the band gap is extremely small. The parametric evolution of angular momentum at $K$ point shows consistent trend with that of Berry curvature. The valley-dependent chirality in a honeycomb lattice and a triangular lattice connect smoothly. Practically, a triangular-hole triangular-lattice PhC serves as an excellent valley PhC which offers evident valley-dependent chirality and large band gap. We envisage this work to provide a clear picture of the valley-photonic properties for various valley PhC designs and raise more discussions in the nature of valley topology.

\section*{Appendix A}

The Berry curvature can also be calculated from the effective Hamiltonian around the K point.
To derive the effective Hamiltonian around \(\textbf{k}_{0}\), we expand the wavefunction by using the wavefunction at \(\textbf{k}_{0}\)\cite{PhysRevB.86.035141,valley5_Wu2017_s0}
\begin{equation}
	\textbf{u}_{n\textbf{k}}(\textbf{r}) = \sum_{j}C_{nj}\textbf{u}_{j\textbf{k}_{0}}(\textbf{r}),
	\label{eq:expansion}
\end{equation}
where $C_{nj}$ is the expansion coefficient.
Inserting Eq. (\ref{eq:expansion}) into Eq. (\ref{eq:Maxwell}) yields
\begin{equation}
	\sum_{j} \biggl[ (\omega_{j\textbf{k}_{0}}-\omega_{n\textbf{k}})B + c
	\begin{pmatrix}
		0 && -\delta \textbf{k} \times
		\\
		\delta \textbf{k} \times && 0
	\end{pmatrix}
	\biggr]\textbf{u}_{j\textbf{k}_{0}} = 0,
\end{equation}
where $\delta\textbf{k}=\textbf{k}-\textbf{k}_{0}$.
Multiplying \(\textbf{u}^{\dagger}_{i\textbf{k}_{0}}\) and integrating over the unit cell, we derive the eigenvalue equation
\begin{gather}
	\sum_{j}[H_{ij}(\delta \textbf{k}) - \omega_{n\textbf{k}}]C_{nj}=0,
	\\
	H_{ij} = \omega_{i\textbf{k}_{0}}\delta_{ij} + \textbf{p}_{ij}\cdot\delta\textbf{k},
	\\
	\textbf{p}_{ij} = \frac{1}{4\sqrt{U_{i\textbf{k}_{0}}U_{j\textbf{k}_{0}}}}\int_{\text{unit cell}} d^{3}r [\textbf{E}^{*}_{i\textbf{k}_{0}}\times\textbf{H}_{j\textbf{k}_{0}} - \textbf{H}^{*}_{i\text{k}_{0}}\times\textbf{E}_{j\textbf{k}_{0}}].
\end{gather}

The effective Hamiltonian $H$ constructed by three bands is explicitly written by
\begin{equation}
	H =
	\begin{pmatrix}
		\omega_{1} + \textbf{p}_{11}\cdot\delta\textbf{k} && \textbf{p}_{12}\cdot\delta\textbf{k} && \textbf{p}_{31}^{*}\cdot\delta\textbf{k}
		\\
		\textbf{p}_{12}^{*}\cdot\delta\textbf{k} && \omega_{2}+\textbf{p}_{22}\cdot\delta\textbf{k} && \textbf{p}_{23}\cdot\delta\textbf{k}
		\\
		\textbf{p}_{31}\cdot\delta\textbf{k} && \textbf{p}_{23}^{*}\cdot\delta\textbf{k} && \omega_{3}+\textbf{p}_{33}\cdot\delta\textbf{k}
	\end{pmatrix}.
\end{equation}
\begin{figure}[hbt!]
	\centering
	\includegraphics[width=0.5\textwidth]{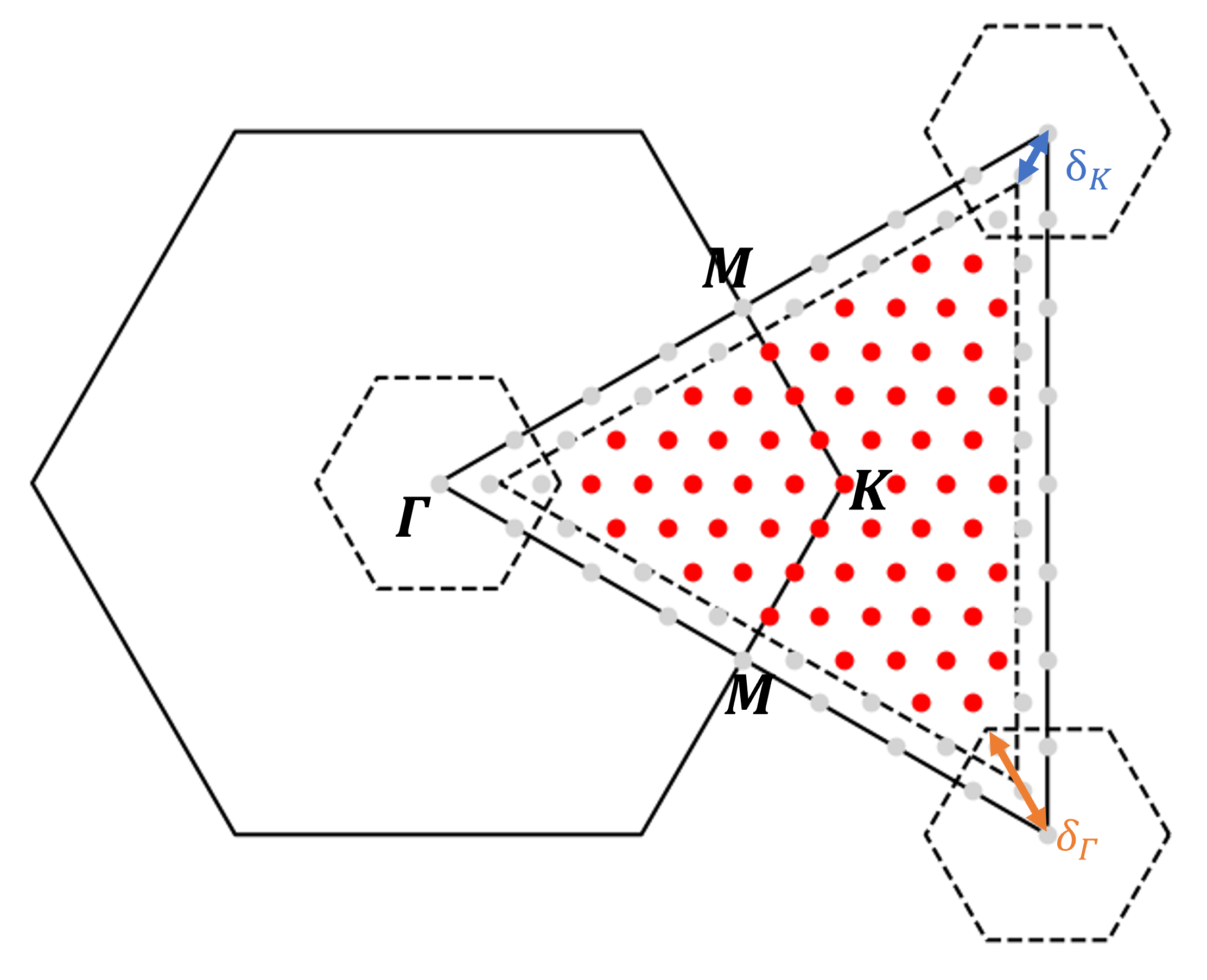}
	\caption{Illustrative scheme of the truncation in Berry curvature calculation. The red and gray dots represent the points in the wavevector space. The solid hexagon shows the first Brillouin zone. The solid triangle shows the half Brillouin zone centered at $K$ point. The dashed triangle shows a shrunken half Brillouin zone, with its radius reduced by $\delta_K$. The area outside the shrunken HBZ is excluded in calculation. The three dashed hexagons centered at $\Gamma$ points indicate another area to be truncated, with a radius of $\delta_{\Gamma}$. The truncated region is shown in gray and the calculated region is shown in red dots. The truncation ratio is exaggerated for visual clarity.  
	}
\end{figure}

If we take the eigenfunction of the three-fold rotational operation as the basis, the three-fold rotational symmetry requires \(\textbf{p}_{ii}=0\) and \(p_{y,ij}=-ip_{x,ij}\)\cite{PhysRevB.86.035141,valley5_Wu2017_s0}.
Replacing \(p_{x,ij}\) with \(a_{i}\), we finally derive the effective Hamiltonian describing the three band around the K point
\begin{equation}
	H =
	\begin{pmatrix}
		\omega_{1} && a_{1}\delta k_{-} && a_{3}^{*}\delta k_{+}
		\\
		a_{1}^{*}\delta k_{+} && \omega_{2} && a_{2}\delta k_{-}
		\\
		a_{3}\delta k_{-} && a_{2}\delta k_{+} && \omega_{3}
	\end{pmatrix},
\end{equation}
where $\delta k_{\pm}=\delta k_{x}\pm\delta k_{-}$ and $a_{i}$ are complex constants. 
The Berry curvature around the K point is calculated from the effective Hamiltonian by\cite{Xiao.D-valley}
\begin{equation}
	\Omega_{n\textbf{k}}=-2\text{Im}\sum_{m\neq n} \frac{\langle \textbf{u}_{n\textbf{k}_{0}} \lvert \frac{\partial H}{\partial k_{x}} \rvert \textbf{u}_{m\textbf{k}_{0}} \rangle \langle \textbf{u}_{m\textbf{k}_{0}} \lvert \frac{\partial H}{\partial k_{y}} \rvert \textbf{u}_{n\textbf{k}_{0}} \rangle}{(\omega_{n\textbf{k}_{0}}-\omega_{m\textbf{k}_{0}})^2}.
\end{equation}

The total Berry curvature of all three bands are conserved for any $\textbf{k}$:
\begin{equation}
	\sum_{n}^{3}\Omega_{n\textbf{k}}=0
\end{equation}

\section*{Appendix B}
\begin{figure*}[hbt!]
	\centering
	\includegraphics[width=1.0\textwidth]{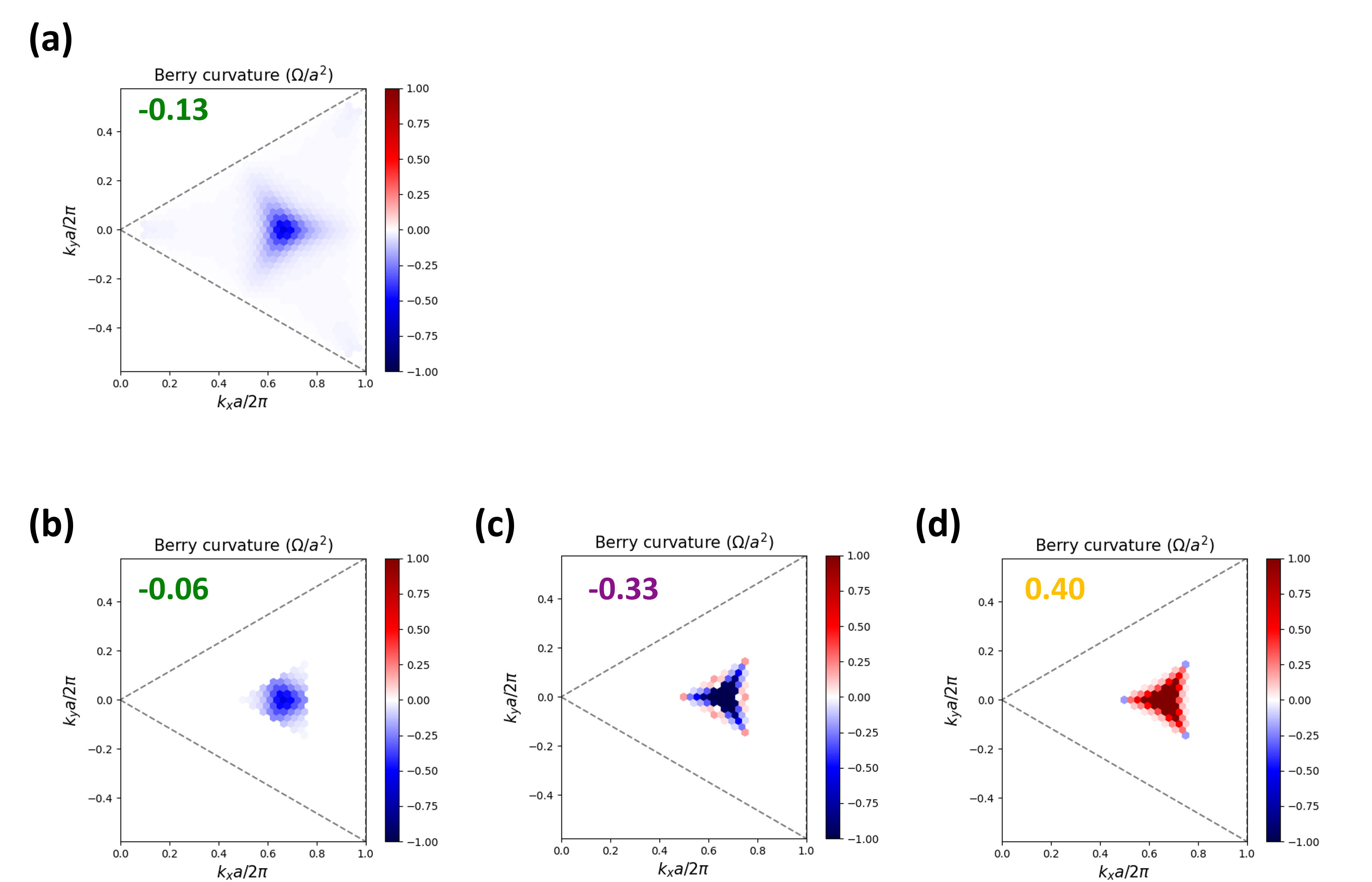}
	\caption{Berry curvature distribution of the $f_c=0.4$ Cir-HPC. The dashed triangles show the half-Brillouin zone centered at K point. The calculated valley Chern number are shown at the left top in each plot.}
\end{figure*}
In this paper we consider the Berry curvature of bands without degeneracies. The Berry curvature is not well defined at degeneracies. Numerical calculation of a precise Berry curvature around a degeneracy is also challenging. Therefore, we only focus on the wavevector region where the first three TE bands are not degenerate. The degeneracy is lifted at $K$ point as long as the $C_{2z}$ symmetry is broken. However, degeneracies are usually inevitable at $\Gamma$ point, especially for band-1 and band-3. Accidental degeneracies may also occur as is the case for band-3 and -4 in the $f_r=0.4$ TPC (Fig.2(c))and band-2 and -3 in the $f_c=0.4$ HPC (Fig.2(d)), which poses some difficulties in the calculation of Berry curvature in the whole HBZ. Therefore, we truncate the calculation region as shown in Fig.7. The solid hexagon shows the FBZ. The solid triangle shows the K-HBZ. The dashed triangle shows a shrunken K-HBZ, whose radius is smaller than the original one by $\delta_K$. The region outside the shrunken K-HBZ will not be calculated. The three dashed hexagons centered at $\Gamma$ points have radius of $\delta_{\Gamma}$. These regions are also excluded in the Berry curvature calculation. Since the focus of this work is the valley-dependent physics defined around $K/K'$ points, this truncation will not affect our main results and conclusion. We also keep a minimum amount of truncation to give a whole picture of the Berry curvature distribution in the wavevector space. For the three PhCs investigated in section 5.Berry curvatures in half-Brillouin zone, we set the $\delta_K$ and $\delta_{\Gamma}$ as follows:

$\delta_K=0.05\frac{2\pi}{a}$, $\delta_{\Gamma}=0.1\frac{2\pi}{a}$ for $f_t=0.8$ HPC.

$\delta_K=0.1\frac{2\pi}{a}$, $\delta_{\Gamma}=0.1\frac{2\pi}{a}$ for $f_r=0$ TPC.
 
$\delta_K=0.05\frac{2\pi}{a}$, $\delta_{\Gamma}=0.13\frac{2\pi}{a}$ for $f_r=0.4$ TPC. 
  
The grid constant (interval) of red dots is $\frac{1}{12}\frac{2\pi}{a}$ in Fig.7. The actual calculation is conducted with a grid constant of $\frac{1}{36}\frac{2\pi}{a}$.

We have not shown the Berry curvature distribution in the main manuscript for the $f_c=0.4$ Cir-HPC because the accidental degeneracies between band-2 and -3 have made the calculation difficult. However, one can calculate the Berry curvatures at vicinity of $K$ point (Fig.8 (b-c)). The truncation setting is $\delta_K=0.5\frac{2\pi}{a}$, $\delta_{\Gamma}=0.1\frac{2\pi}{a}$. Near $K$ point, $\Omega_2$ and $\Omega_3$ have approximately opposite values. $\Omega_1$ is relatively small. The integrated Berry curvature (or a incomplete valley Chern number) in the calculate region is -0.06,-0.33 and 0.40 for the first three bands. The full distribution of the first band of $f_c=0.4$ Cir-HPC (which has no degeneracies except at $\Gamma$ point) is shown in Fig.8 (a), with $\delta_K=0.05\frac{2\pi}{a}$, $\delta_{\Gamma}=0.1\frac{2\pi}{a}$. Even in this region the integrated $\Omega_1/a^2$ is small, only 0.13. This is consistent with the analytical prediction in $k\cdot p$ method.

\section*{data availability}
	Data supporting the findings in this paper is available at \doi{10.6084/m9.figshare.31076839}

\begin{acknowledgments}
	The authors thank Takahiro Uemura for assistance in calculation and clarifying key concepts. This research was funded by Japan Society for the Promotion of Science (JP20H05641; KAKENHI 24K01377, 24H02232 and 24H00400).
\end{acknowledgments}


\bibliography{reference1}

\end{document}